%
%
%
%
%
%
%
%
%
%
%
%
%
%
%
\input phyzzx
\input epsf
\newcount\figno
\figno=0
\def\fig#1#2#3{
\par\begingroup\parindent=0pt\leftskip=1cm\rightskip=1cm\parindent=0pt
\baselineskip=11pt
\global\advance\figno by 1
\midinsert
\epsfxsize=#3
\centerline{\epsfbox{#2}}
\vskip 12pt
{\bf Fig. \the\figno:} #1\par
\endinsert\endgroup\par
}

%
%
\catcode`\@=11 
\def\papersize{\hsize=40pc \vsize=53pc \hoffset=0pc \voffset=1pc
   \advance\hoffset by\HOFFSET \advance\voffset by\VOFFSET
   \pagebottomfiller=0pc
   \skip\footins=\bigskipamount \normalspace }
\catcode`\@=12 

\def\sin{\mathop{\rm sin}\nolimits}
\def\der{\partial}
\def\Im{\mathop{\rm Im}\nolimits}

\def\Gslash{\, /\hskip-0.25cm G}
\def\smallGslash{\, /\hskip-0.20cm G}
\def\Cslash{\, /\hskip-0.25cm C}
\def\aslash{\, /\hskip-0.21cm a}
\def\kslash{\, /\hskip-0.22cm k}
\def\wslash{\, /\hskip-0.23cm w}
\def\lambdaslash{\, /\hskip-0.24cm \lambda}

\papers


\vsize=23.cm
\hsize=15.cm

\tolerance=500000
\overfullrule=0pt

\Pubnum={LPTENS-97/19 \cr
{\tt cond-mat/9705044} \cr
April 1997}

\date={}
\pubtype={}
\titlepage
\title{\bf Universality of correlation functions of \break
hermitian random matrices in an external field\break
}
\author{P.~Zinn-Justin}
\vskip 1.cm
\address{
Laboratoire de Physique Th\'eorique de l'Ecole
Normale Sup\'erieure
\foot{{\rm unit\'e propre du CNRS, associ\'ee \`a l'Ecole Normale
Sup\'erieure et l'Universit\'e Paris-Sud.}}
 \break 24 rue Lhomond, 75231
Paris Cedex 05, France\break
{\tt pzinn@physique.ens.fr}
} 

\vskip 1.cm
\abstract{The behavior of correlation functions is studied
in a class of matrix models characterized
by a measure $\exp(-S)$ containing a potential term
and an external source term: $S=N\tr(V(M)-MA)$.
In the large $N$ limit,
the short-distance behavior is found to be identical to the one obtained in previously studied
matrix models, thus extending the universality of the level-spacing distribution.
The calculation of correlation functions involves (finite $N$) determinant formulae,
reducing the problem to the large $N$ asymptotic analysis of a single kernel $K$.
This is performed by an appropriate matrix integral formulation of $K$. 
Multi-matrix generalizations of these results are discussed.



\endpage
\pagenumber=1

\REF\WI{E.~P.~Wigner, {\it Proc. Cambridge Philos. Soc.} 47 (1951) 790,
and other papers reprinted in C.~E.~Porter, {\it Statistical theories of spectra: fluctuations} (Academic Press,
New York, 1991).}
\REF\ME{M.~L.~Mehta, {\it Random matrices}, 2nd ed. (Academic Press,
New York 1991).}
\REF\DY{F.~J.~Dyson, {\it J. Math. Phys.} 13 90 (1972).}
\REF\IZ{Harish~Chandra, {\it Amer. J. Math.} 79 (1957) 87-120. \hfill\break
C.~Itzykson and J.-B.~Zuber, {\it J. Math. Phys.} 21 (1980) 411.}
\REF\PA{L.~A.~Pastur, {\it Theor. Math. Phys. (USSR)} 10 (1972) 67.}
\REF\KAZ{V.~A.~Kazakov, {\it Nucl. Phys.} B (Proc. Suppl.) 4 (1988) 93.}
\REF\AMB{J.~Ambjorn and Yu.~M.~Makeenko, {\it Mod. Phys. Lett.}
A 5 (1990) 1753.}
\REF\GRN{D.~J.~Gross and M.~J.~Newman, {\it Phys. Lett.} B 266 (1991) 291.}
\REF\BZJ{E.~Br\'ezin and J.~Zinn-Justin, {\it Phys. Lett.} B 288 (1992) 54.
\hfill\break
S.~Higuchi, C.~Itoi, S.~Nishigaki, N.~Sakai, {\it Nucl. Phys.} B 318 (1993) 63.}
\REF\ZJE{B.~Eynard and J.~Zinn-Justin, {\it Nucl. Phys.} B 386 (1992) 558.}
\REF\BZ{E.~Br\'ezin and A.~Zee, {\it Nucl. Phys.} B 402 (1993) 613.}
\REF\MEL{P.~A.~Mello, {\it Theory of random matrices:
spectral statistics and scattering problems}
in {\it Mesoscopic quantum physics}, Les Houches Session
LXI, E.~Akkermans, G.~Montambaux, J.-L.~Pichard, and J.~Zinn-Justin eds.
(North-Holland, 1994), and references therein.}
\REF\MAT{A.~Matytsin, {\it Nucl. Phys.} B 411 (1994) 805.}
\REF\BEEN{C.~W.~J.~Beenakker, {\it Nucl. Phys.} B 422 (1994) 515.}
\REF\EY{B.~Eynard, {\it Gravitation quantique bidimensionnelle et matrices
al\'eatoires}, th\`ese de doctorat de l'Universit\'e Paris 6 (1995).}
\REF\KSW{V.~A.~Kazakov, M.~Staudacher and T.~Wynter. {\it Commun. Math. Phys.} 177 (1996), 451;
\hfill\break
{\it Commun. Math. Phys.} 179 (1996) 235.}
\REF\BH{E.~Br\'ezin and S.~Hikami, {\it Nucl. Phys.} B 479 (1996) 697.\hfill\break
E.~Br\'ezin and S.~Hikami, preprint cond-mat/9608116.\hfill\break
E.~ Br\'ezin and S.~Hikami, preprint cond-mat/9702213.}
\REF\PZJ{P.~Zinn-Justin, preprint cond-mat/9703033.}
\REF\DEO{N.~Deo, preprint cond-mat/9703136.}

\chapter{Introduction.}
More than four decades ago, Wigner [\WI] suggested to study the distribution
of energy levels of complex systems using random matrices.
In this approach, one would like to
characterize the structure of the energy levels, i.e. the eigenvalues
of the Hamiltonian, the latter being considered as a large matrix
with random entries.
It is now known that many statistical properties of spectra
of true physical systems are indeed well described by those of random matrices
(cf [\MEL] for a review): it is therefore important to understand how much these
spectral properties depend on the particular matrix ensemble chosen, i.e. determine
universality classes of matrix ensembles.

For technical reasons,
we shall consider here ensembles of hermitian matrices, which
correspond to systems without time-reversal invariance.
The main quantities of interest are the probability distributions $\rho_n$ 
of the eigenvalues: if $M$ is a random hermitian $N\times N$
matrix, we define $\rho_n(\lambda_1,\ldots,\lambda_n)$ to be
the density of probability that $M$ has $(\lambda_1,\ldots,\lambda_n)$
among its $N$ eigenvalues, with the normalization convention that:
$\int\prod_{i=1}^{n} d\lambda_i\, \rho_n(\lambda_1,\ldots,\lambda_n)=1.$
To connect with correlations functions of the model,
we define (following [\ME]):
$$R_n(\lambda_1,\lambda_2,\ldots,\lambda_n)
\equiv\left<\prod_{i=1}^n\tr\delta(M-\lambda_i)\right>.
\eqn\cor$$
Then one has $R_n(\lambda_1,\ldots,\lambda_n)={N!\over (N-n)!}\rho_n(\lambda_1,\ldots,
\lambda_n)$ for distinct $\lambda_i$ ($\delta$ functions appear for coinciding eigenvalues).
This means that correlation functions of any $U(N)$-invariant quantities
(i.e. functions of the eigenvalues only)
can be computed using the $\rho_n$.

We shall now study a class of models in
which one can express the
functions $\rho_n$ in terms of a single kernel $K(\lambda,\mu)$; if we again assume
the $\lambda_i$ all distinct, the corresponding relation for $R_n$ will be
$$R_n(\lambda_1,\lambda_1,\ldots,\lambda_n)=
\det (K(\lambda_i,\lambda_j))
_{i,j=1\ldots n}.\eqn\detform$$
These formulae are exact at finite $N$.

For example if we also define
$$R_n^{(c)}(\lambda_1,\lambda_2,\ldots,\lambda_n)\equiv
\left<\prod_{i=1}^n\tr\delta(M-\lambda_i)\right>_c \eqn\concor$$
the connected correlation functions, this implies that
$$R_2^{(c)}(\lambda,\mu)=-K(\lambda,\mu)K(\mu,\lambda).\eqn\concorb$$
In these models,
the study of the distribution of eigenvalues reduces
to the analysis of this kernel; in particular the large $N$ limit
of $K$ should allow to compute all correlation
functions in this limit
and find the different ``universal'' behaviors that can
arise:

$\star$ The long distance behavior. As it is known that the kernel fluctuates wildly
on intervals of size $\sim 1/N$, one must first average the kernel
to suppress the oscillations on this scale and obtain a sensible long
distance behavior.

$\star$ The short distance behavior. This is the region $\lambda-\mu\sim 1/N$,
in which the fast oscillations mentioned above are relevant.

The long distance behavior can be studied by various
standard large $N$ techniques
[\AMB,\BZ,\BEEN], so we shall concentrate here on
the short distance behavior. Usually one characterizes this behavior by
introducing the level-spacing distribution $P(s)$, $s=N(\lambda-\mu)$
($P(s)$ a priori also depends on $\lambda$ or $\mu$).
In the large $N$ limit,
$P(s)$ can be simply related to the asymptotic form $\hat{K}$ of the
kernel on $[\lambda,\mu]$:
$$P(s)={d^2\over ds^2} \det\left[1-{1\over N}\hat{K}\right]\eqn\lvlspc$$
where $\det$ is the Fredholm determinant. Thus, in the short distance region,
the universality of $K$ implies the universality of the level
spacing.

In section 2 we first consider the case of a matrix model,
where the measure consists
of a simple potential term $\tr V(M)$. Expressions of the kernel $K$
in terms of orthogonal polynomials have been known for a long time [\ME].
One can then proceed to derive a short distance universal behavior
of the kernel [\BZ,\EY]:
$$K(\lambda,\mu)\sim {\sin x\over x},\qquad x\sim N(\lambda-\mu).\eqn\sdu$$
Here we shall rewrite $K$, and rederive its large $N$ asymptotic form,
using a new
method which does not make use of  orthogonal polynomials, keeping in mind
that we are ultimately interested in the more difficult case of the model
of matrices coupled to an external field.

The first hint that the latter model could possess the same short distance
universal behavior appeared in a series of papers [\BH] by Brezin and Hikami,
who showed that formula $\detform$
can be generalized to the
gaussian ensemble with an external field, that is for the measure 
$\exp \left(-{N\over 2} \tr M^2 + N\tr MA\right) d^{N^2}\! M$.
This measure can be interpreted by saying that M is the sum
of a fixed Hamiltonian $A$ and a random gaussian part $M-A$: the constant part
$A$ breaks the $U(N)$ invariance of the model
so neither orthogonal polynomials (like in
the standard one-matrix model) nor biorthogonal polynomials (like in the
two-matrix model) are of any use. Still, one can define a kernel in this
model, which has the same short distance behavior $\sdu$.

Then in [\PZJ] the more general case of a measure
of the type $\exp(N\tr(-V(M)+MA))$ was investigated. The motivation of this
measure was to study a random Hamiltonian which contains a not necessarily
gaussian potential, and an external source term which breaks the
$U(N)$-invariance: could the latter symmetry be somehow related to the
universality of the level spacing distribution ?

To answer this question,
we shall again define in section 3 a kernel $K$ such that equation $\detform$
holds. Then, just as in section 2, we shall rewrite
$K$ as a matrix integral, allowing to take the $N\rightarrow\infty$ limit,
and find the short distance universal behavior.

The methods used here are very general, and in particular, they can
be successfully applied to multi-matrix models; to show this we study
in section 4 
a model of a chain of matrices, with or without an external field at the
end of the chain. We give without proof, and in analogy with the
one-matrix model, expressions for the kernel $K$, which exhibit the
same short distance universality.

Finally, appendix 1 describes in detail the analytic structure
of the functions involved in the large $N$ limit,
and appendix 2 shows the connection
between the formalism used here and large $N$ character formulae.

\vfill\eject

\chapter{The $U(N)$-invariant case.}
\section{Definition of the model and of the kernel.}
Let us consider an ensemble
of random hermitian $N\times N$ matrices with the measure
$$Z^{-1} \exp \left(-N\tr V(M)\right) d^{N^2}\! M \eqn\meas$$
where $V$ is a polynomial and $Z$ the partition function. An important
remark is that this is {\it not} the most general $U(N)$-invariant measure
(one could have products of traces of functions of $M$).

A classical result [\ME] expresses
the distribution law $\rho_n$ of $n$ eigenvalues ($1\le n \le N$) of $M$
in terms of the kernel
$$K(\lambda,\mu)=\sum_{k=0}^{N-1} F_k(\lambda) F_k(\mu).\eqn\stdker$$
Here $F_i$ is the orthonormal function associated to the usual orthogonal
polynomial $P_i(\lambda)=\lambda^i+\cdots$:
$$\eqalign{
&F_i(\lambda)=h_i^{-1/2} P_i(\lambda) e^{-{N\over 2} V(\lambda)}\cr
&\int d\lambda e^{-N V(\lambda)} P_i(\lambda) P_j(\lambda) = h_i \delta_{ij}.\cr
}\eqn\orth$$
(see [\EY] for a review of orthogonal polynomials in matrix models).
Let us briefly rederive this result.
As the measure $\meas$ only depends on the eigenvalues of $M$, the integration
over the angular variables is trivial and one finds:
$$\rho_N(\lambda_1,\lambda_2,\ldots,\lambda_N)=Z^{-1} \Delta^2(\lambda_i)
e^{-N \sum_{i=1}^N V(\lambda_i)}.\eqn\rhoN$$
The Van der Monde determinant $\Delta(\lambda_i)=\det (\lambda_i{}^j)
_{1\le i\le N,0\le j\le N-1}$ can be rewritten in terms of the orthogonal polynomials:
$$\rho_N(\lambda_1,\lambda_2,\ldots,\lambda_N)=Z^{-1} \det (P_k(\lambda_i))
_{1\le i\le N,0\le k\le N-1} \det (P_k(\lambda_j))_{1\le j\le N,
0\le k\le N-1}
e^{-N \sum_{i=1}^N V(\lambda_i)}.\eqn\rhoNb$$
One can now easily compute $Z=N! \prod_{i=0}^{N-1} h_i$ by integrating over
all $\lambda_i$. Combining the two determinants, we finally obtain:
$$\rho_N(\lambda_1,\lambda_2,\ldots,\lambda_N)= {1\over N!}
\det (K(\lambda_i,\lambda_j))
_{i,j=1\ldots N}.\eqn\rhoNc$$
The kernel $K$ has the following properties:
$$\left\{\eqalign{
K(\lambda,\mu)&=K(\mu,\lambda)\cr
[K\star K](\lambda,\nu) &\equiv \int d\mu \, K(\lambda,\mu) K(\mu,\nu)
= K(\lambda,\nu) \cr
}\right.\eqn\kerprop$$
i.e. it is the orthogonal projector on the subspace spanned by the $F_k$,
$0\le k\le N-1$. Using the property $K\star K=K$ and noting that
$$\rho_n(\lambda_1,\lambda_2,\ldots,\lambda_n)=\int d\lambda_{n+1}
\rho_{n+1}(\lambda_1,\lambda_2,\ldots,\lambda_{n+1}),\eqn\induc$$
one can then show inductively that
$$\rho_n(\lambda_1,\lambda_2,\ldots,\lambda_n)={(N-n)!\over N!}
\det (K(\lambda_i,\lambda_j))_{i,j=1\ldots n}.\eqn\detformb$$
for any $n\le N$. This is equivalent to formula $\detform$.

\section{Matrix integral formulation of the kernel.}
We shall now rewrite $K(\lambda,\mu)$ as a matrix integral:
$$K(\lambda,\mu)=Z^{-1} e^{-{N\over 2}(V(\lambda)+V(\mu))}
\int d^{(N-1)^2}\! M \det(\lambda-M) \det(\mu-M) \exp (-N\tr V(M)). \eqn\kermat$$
It is easy to check this formula by going over to eigenvalue variables:
$$\eqalign{
&\int d^{(N-1)^2}\! M \det(\lambda-M) \det(\mu-M) \exp (-N\tr V(M))\cr
&=\int \prod_{i=1}^{N-1} \left( d\lambda_i (\lambda-\lambda_i) (\mu-\lambda_i)\right)
\Delta^2(\lambda_i)_{1\le i\le N-1}
e^{-N\sum_{i=1}^{N-1} V(\lambda_i)}\cr
&=\int \prod_{i=1}^{N-1} d\lambda_i
\Delta(\lambda_i)_{1\le i\le N, \lambda_N\equiv\lambda}
\Delta(\lambda_i)_{1\le i\le N, \lambda_N\equiv\mu}
e^{-N\sum_{i=1}^{N-1} V(\lambda_i)}\cr
&=\sum_{\sigma,\sigma'\in {\cal S}_N} (-1)^\sigma (-1)^{\sigma'}
P_{\sigma(N)}(\lambda) P_{\sigma'(N)}(\mu)
\prod_{i=1}^{N-1}\left[ \int dz P_{\sigma(i)}(z) P_{\sigma'(i)}(z)
e^{-NV(z)}\right]\cr
}\eqn\kerproof$$

This is non-zero when $\sigma=\sigma'$ and we find as expected:
$$K(\lambda,\mu)=e^{-{N\over 2}(V(\lambda)+V(\mu))}
\sum_{k=0}^{N-1} h_k^{-1} P_k(\lambda) P_k(\mu).\eqn\kerconc$$
Note that equation $\kermat$ does not involve orthogonal polynomials, and that is why
we shall be able to generalize it to the non $U(N)$-invariant
model of matrices coupled to an external field,
for which the orthogonal polynomials formalism is not available.

\section{Large N asymptotics of the kernel.}
Formula $\kermat$ allows to compute asymptotics of $K(\lambda,\mu)$
as $N\rightarrow\infty$. It relates $K(\lambda,\mu)$
to the partition function $Z(\lambda,\mu)$ of a matrix
model with the measure 
$\exp(\tr\log(\lambda-M)+\tr\log(\mu-M)-N\tr V(M))
d^{(N-1)^2}\! M$:
$$Z(\lambda,\mu)\equiv\int d^{(N-1)^2}\! M \det(\lambda-M) \det(\mu-M) \exp (-N\tr V(M)).\eqn\modmeas$$
Rather than directly applying the saddle point method to this
expression, it is easier to write differential equations for $Z$.
Indeed one has:
$$\left\{ \eqalign{
{\der\over\der\lambda} \log Z(\lambda,\mu)&=
(N-1) G_{\lambda,\mu}(\lambda)\cr
{\der\over\der\mu} \log Z(\lambda,\mu)&=
(N-1) G_{\lambda,\mu}(\mu) \cr}\right.\eqn\eqdif$$
where $G_{\lambda,\mu}$ is the resolvent of this model:
$$G_{\lambda,\mu}(z)={1\over N-1} \left<\tr{1\over z-M}\right>\eqn\res$$
which depends on $\lambda$ and $\mu$ through the $\tr\log(\lambda-M)$
and $\tr\log(\mu-M)$ terms in the action.
Note that the factor $N-1$ in front of $G_{\lambda,\mu}$
in $\eqdif$ forces us to compute $G$ up to $1/N$ corrections. 

Let us now find a saddle point for the eigenvalues.
In the large $N$ limit, we shall suppose that they fill
a single interval $[\alpha,\beta]$; then
$G_{\lambda,\mu}(z)$ becomes an analytic function of $z$ with a single
cut on $[\alpha,\beta]$:
$$G_{\lambda,\mu}(z\pm i0)=\Gslash_{\lambda,\mu}(z)
\pm i\pi\rho_{\lambda,\mu}(z)\eqn\cut$$
where $\rho_{\lambda,\mu}(z)$ is the density of eigenvalues at $z$.
As a definition $\Gslash(z)\equiv
{1\over 2}(G(z+i0)+G(z-i0))$ for any function $G$.
The saddle point equation can now be written:
$$2(N-1)\Gslash_{\lambda,\mu}(z)-NV'(z)+{1\over z-\lambda}+{1\over z-\mu} = 0
\qquad\forall z\in [a,b].\eqn\spe$$
At leading order in $N$, this equation is just the usual saddle point
equation $2\Gslash (z)=V'(z)$ for the resolvent $G$ of
the original matrix model we have started
from. Therefore we can write:
$$G_{\lambda,\mu}=G+{1\over N}(C_G+C_\lambda+C_\mu),\eqn\defcor$$
where we have introduced three $1/N$ corrections to the leading behavior of
$G_{\lambda,\mu}$ corresponding to the three corrective terms in the saddle
point equation $\spe$. Following [\ZJE,\EY] we deduce these corrections from
their analytic properties. Both $G(z)$ and $G_{\lambda,\mu}(z)$ behave as
$1/z$ as $z\rightarrow\infty$, so $C_G$, $C_\lambda$ and $C_\mu$ are
$O(1/z^2)$. Furthermore they satisfy
$$\left\{\eqalign{
\Cslash_G(z)&=\Gslash(z)\cr
\Cslash_\lambda(z)&={1\over 2(\lambda-z)}\cr
\Cslash_\mu(z)&={1\over 2(\mu-z)}\cr
}\right.
\eqn\slacor$$
and are regular for all $z$ except near $\alpha$ (resp. $\beta$) where they
should behave as $1/\sqrt{z-\alpha}$ (resp. $1/\sqrt{z-\beta}$). This
determines them entirely:
$$\left\{\eqalign{
C_G(z)&=G(z)-{1\over\sqrt{(z-\alpha)(z-\beta)}}\cr
C_\lambda(z)&={1\over 2} {1\over\sqrt{(z-\alpha)(z-\beta)}}
\left(1-{\sqrt{(z-\alpha)(z-\beta)}-\sqrt{(\lambda-\alpha)(\lambda-\beta)}
\over z-\lambda}\right)\cr
C_\mu(z)&={1\over 2} {1\over\sqrt{(z-\alpha)(z-\beta)}}
\left(1-{\sqrt{(z-\alpha)(z-\beta)}-\sqrt{(\mu-\alpha)(\mu-\beta)}
\over z-\mu}\right).\cr
}\right.\eqn\cor$$
The right hand side of $\eqdif$ can now be written:
$$(N-1) G_{\lambda,\mu}(\lambda)
=N G(\lambda)-{1\over 2} {d\over d\lambda} \log\sqrt{(\lambda-\alpha)(\lambda-
\beta)} -{1\over 2} {1\over\lambda-\mu}
\left(1-\sqrt{(\mu-\alpha)(\mu-\beta)\over (\lambda-\alpha)(\lambda-\beta)}
\right)\eqn\eqdifb$$
and a similar equation for $(N-1) G_{\lambda,\mu}(\mu)$.
For $\lambda\in [\alpha,\beta]$,
an ambiguity in $\eqdifb$ must be resolved: $G_{\lambda,\mu}$, like
$G$, has a cut on $[\alpha,\beta]$;
so we must choose $\lambda$ slightly above or
below the real axis to determine the right hand side of $\eqdifb$:
$$\eqalign{
(N-1) G_{\lambda,\mu}(\lambda\pm i0)
&={N\over 2}V'(\lambda)\pm Ni\pi\rho(\lambda)
-{1\over 2} {d\over d\lambda} \log\sqrt{(\lambda-\alpha)(\beta-\lambda)}\cr
&-{1\over 2} {1\over\lambda-\mu} \left(1-{\sqrt{(\mu-\alpha)(\mu-\beta)}
\over \pm i\sqrt{(\lambda-\alpha)(\beta-\lambda)}}\right).\cr
}\eqn\ambi$$
($\rho(\lambda)\equiv\rho_1(\lambda)$).
This ambiguity, which appears only at $N=\infty$, means that there are several
saddle points which we must all take into account. The same problem appears
when $\mu$ gets close to the cut, so there is a total
of 4 saddle points $(\epsilon,\epsilon')$ ($\epsilon$, $\epsilon'=\pm$)
corresponding to the locations of $\lambda$ and $\mu$ with respect to
the cut $[\alpha,\beta]$.

Finally we can write differential equations for $K(\lambda,\mu)$:
$$\eqalign{
{\der\over\der\lambda} \log K_{(\epsilon,\epsilon')} &= \epsilon Ni\pi\rho(\lambda)
-{1\over 2} {d\over d\lambda} \log\sqrt{(\lambda-\alpha)(\beta-\lambda)}
-{1\over 2} {1\over\lambda-\mu} \left(1-{\epsilon'\sqrt{(\mu-\alpha)(\beta-\mu)}
\over \epsilon \sqrt{(\lambda-\alpha)(\beta-\lambda)}}\right)\cr
{\der\over\der\mu} \log K_{(\epsilon,\epsilon')} &= \epsilon' Ni\pi\rho(\mu)
-{1\over 2} {d\over d\mu} \log\sqrt{(\mu-\alpha)(\beta-\mu)}
-{1\over 2} {1\over\mu-\lambda} \left(1-{\epsilon\sqrt{(\lambda-\alpha)(\beta-\lambda)}
\over \epsilon' \sqrt{(\mu-\alpha)(\beta-\mu)}}\right).\cr
}\eqn\eqdifc$$
We introduce the function $\varphi(z)$ which satisfies
$z={1\over 2}(\alpha+\beta)-{1\over 2}(\beta-\alpha)\cos\varphi(z)$ and ${1\over 2}(\beta-\alpha)
\sin\varphi(z)=\sqrt{(z-\alpha)(\beta-z)}$. Noting then that
$${d\over d\lambda} \log\sin\left({\epsilon\varphi(\lambda)-\epsilon'\varphi(\mu)\over 2}\right)
=\epsilon{1\over\sqrt{(\lambda-\alpha)(\beta-\lambda)}}
{\epsilon\sqrt{(\lambda-\alpha)(\beta-\lambda)}+\epsilon'\sqrt{(\mu-\alpha)(\beta-\mu)}
\over \lambda-\mu},\eqn\fun$$
equations $\eqdifc$ can be integrated:
$$K_{(\epsilon,\epsilon')}(\lambda,\mu)=c_{(\epsilon,\epsilon')}
{\sin\left({\epsilon\varphi(\lambda)-\epsilon'\varphi(\mu)\over 2}\right)\over \lambda-\mu}
{1\over\sqrt{\sin\varphi(\lambda)\sin\varphi(\mu)}}
\exp\left( \epsilon iN\pi \int_{\lambda_0}^\lambda \rho(z) dz
+\epsilon' iN\pi \int_{\lambda_0}^\mu \rho(z) dz \right).\eqn\theker$$
The integration constants
$c_{(\epsilon,\epsilon')}$ satisfy $c_{(\epsilon,\epsilon')}
=c_{(\epsilon',\epsilon)}$ (interchange of $\lambda$ and $\mu$)
and $c_{(-\epsilon,-\epsilon')}=-\bar{c}_{(\epsilon,\epsilon')}$
(complex conjugation). $c_{(\pm,\mp)}$ are independent of the
choice of $\lambda_0$ and are fixed by imposing the normalization
condition $K(\lambda,\lambda)=N\rho(\lambda)$:
we find (cf next section) that $c_{(\pm,\mp)}=1/(2\pi i)$.
$c_{(\pm,\pm)}$ are undetermined; if we assume that one can find
$\lambda_0$ such that $c_{(\pm,\pm)}=\pm 1/2\pi$
(for the case of an
even potential we would have $\lambda_0=0$), then we are left with only
one unknown parameter $\lambda_0$.

We can finally sum the four
function $K_{(\epsilon,\epsilon')}$; we obtain:
$$\eqalign{
K(\lambda,\mu)=&{1\over\pi}{1\over\lambda-\mu}{1\over\sqrt{\sin\varphi(\lambda)
\sin\varphi(\mu)}}\cr
&\left[\sin\left({\varphi(\lambda)+\varphi(\mu)\over 2}\right)
\sin\left(N\pi\int^\lambda_\mu \rho(z) dz \right)\right.\cr
&\left.+\sin\left({\varphi(\lambda)-\varphi(\mu)\over 2}\right)
\cos\left(N\pi\int_{\lambda_0}^\lambda\rho(z)dz
+N\pi\int_{\lambda_0}^\mu\rho(z)dz\right)\right],\cr
}\eqn\thekerb$$
a formula for $K$ that
is equivalent to the one that was found in [\BZ] by using an ansatz on
the form of orthogonal polynomials (see also [\EY]).

\section{Short distance universal behavior of the kernel.}
Let us now inspect the region where $\lambda-\mu\sim 1/N$, $\alpha<\lambda<\beta$.
It is clear from $\theker$-$\thekerb$ that
the dominant contributions come here from the saddle points $(\pm,\mp)$ (i.e. $\lambda$ and $\mu$
on opposite sides of the cut). Actually, this can already be seen in the equations $\eqdifc$,
which acquire a particularly simple form in this limit:
$$\left\{\eqalign{
{d\over d\lambda} \log K_{(\pm,\mp)} &= \pm Ni\pi\rho(\lambda)
-{1\over\lambda-\mu}\cr
{d\over d\mu} \log K_{(\pm,\mp)} &= \mp Ni\pi\rho(\mu)
-{1\over\mu-\lambda}\cr
}\right. \eqn\eqdifsimp$$
Here we do not need these simplified differential equations, since we can
directly take the limit in $\thekerb$, which yields
$$K(\lambda,\mu)={\sin(N\pi(\lambda-\mu)\rho(\lambda))
\over\pi(\lambda-\mu)}.
\eqn\asymp$$
This is the well-known short distance universal behavior $\sdu$ of the kernel.

\chapter{Generalization to the case of an external field.} 
It was proven in [\PZJ] that
in the case of a general measure with an external field, $\detform$ still holds;
we shall now review this result and write the kernel $K$
in an appropriate way for asymptotic analysis.

\section{Definition of the model and of the kernel.}
Let us consider the measure:
$$Z^{-1} \exp \left(-N \tr V(M)+N \tr MA\right) d^{N^2}\! M \eqn\measb$$
where $V$ is an arbitrary polynomial, and $A={\rm diag}
(a_1,\ldots,a_N)$ can be assumed diagonal. Particular
matrix models of this type appear in several papers [\KAZ,\GRN]. 

One diagonalizes $M$: if $M=\Omega\Lambda\Omega^\dagger$
where $\Lambda={\rm diag}(\lambda_1,\ldots,\lambda_N)$, 
the integral over $\Omega$ is the usual Itzykson--Zuber integral [\IZ]
on the unitary group and we find:
$$\rho_N(\lambda_1,\lambda_2,\ldots,\lambda_N)= Z^{-1} \Delta(\lambda_i)
{\det (\exp (N \lambda_j a_l)) \over \Delta(a_l)}
e^{-N \sum_{i=1}^N V(\lambda_i)}.\eqn\rhoNd$$
$Z$ can now be computed:
$$\eqalign{
Z&=N!{1\over \Delta(a_l)} \int \prod_{i=1}^N d\lambda_i \det(\lambda_i{}^k)
_{1\le i\le N,0\le k\le N-1}
e^{N \sum_{i=1}^N (-V(\lambda_i) + a_i \lambda_i)}\cr
&= {N!\over \Delta(a_l)}\det\left(\int d\lambda\, \lambda^k 
e^{N (-V(\lambda)+a_l \lambda)}\right)
_{0\le k\le N-1,1\le l\le N}\cr
} \eqn\Z$$
Inserting $\Z$ into $\rhoNd$ yields
$$\rho_N(\lambda_1,\lambda_2,\ldots,\lambda_N)={1\over N!}
{\det(\lambda_i{}^k)_{1\le i\le N,0\le k\le N-1}\,
\det(\exp (N a_l \lambda_j))_{1\le j,l\le N}
\over \det\left(\int d\lambda\, \lambda^k e^{N (-V(\lambda)+a_l \lambda)}\right)
_{0\le k\le N-1,1\le l\le N}}
e^{-N \sum_{i=1}^N V(\lambda_i)}.\eqn\longdet$$
The matrix $m_{lk}=\int d\lambda\, \lambda^k \exp N (-V(\lambda)+a_l \lambda)$
possesses an inverse, which we denote by $\alpha_{kl}$; putting
together the three determinants (and the $\exp -N V(\lambda)$ factors)
we finally obtain:
$$\rho_N(\lambda_1,\lambda_2,\ldots,\lambda_N)={1\over N!}
\det(K(\lambda_i,\lambda_j))_{i,j=1\ldots N}\eqn\detN$$
where
$$K(\lambda,\mu)=e^{-{N\over 2} (V(\lambda)+V(\mu))}
\sum_{k=0}^{N-1}\sum_{l=1}^N \alpha_{kl} \lambda^k e^{Na_l\mu}.
\eqn\defK$$
The kernel $K$ satisfies the property:
$$\eqalign{
[K\star K](\lambda,\rho)&=e^{-{N\over 2} (V(\lambda)+V(\rho))}
\sum_{k,k',l,l'} \alpha_{kl} \lambda^k
\left[ \int d\mu\, \mu^{k'} e^{Na_l\mu-NV(\mu)} \right]
\alpha_{k'l'} e^{Na_{l'}\rho}\cr
&=K(\lambda,\rho).\cr
}\eqn\kerpropb$$
Thus, one can follow the same line of reasoning as
in the $U(N)$-invariant case to obtain the determinant formulae
$$\rho_n(\lambda_1,\lambda_2,\ldots,\lambda_n)={(N-n)!\over N!}
\det(K(\lambda_i,\lambda_j))_{i,j=1\ldots n}\eqn\detformc$$
for any $n\le N$.

If we introduce
the polynomials $Q_l$:
$$Q_l(\lambda)=\sum_{k=0}^{N-1} \alpha_{kl} \lambda^k$$
then
$$K(\lambda,\mu)=e^{-{N\over 2} (V(\lambda)+V(\mu))}
\sum_{l=1}^N Q_l(\lambda) e^{Na_l\mu}.$$
The polynomials $Q_l$ are of degree $N-1$, and satisfy the orthogonality
relations:
$$\int d\lambda Q_l(\lambda) e^{N (-V(\lambda)+a_{l'}\lambda)}
=\delta_{ll'},\eqn\orth$$
This proves that $K$ is a non-orthogonal projector on the space spanned
by the $Q_l(\lambda) \exp(-NV(\lambda)/2)$, $1\le l\le N$, which is also
the space spanned by the $F_k$, $0\le k\le N-1$.

\section{Matrix integral formulation of the kernel.}
We shall now guess matrix integral formulae for the polynomials $Q_l$:
$$Q_l(\lambda)=c_l \int d^{(N-1)^2}\! M \det(\lambda-M)
\exp(N\tr(-V(M)+MA^{(l)})).\eqn\matpol$$
Here $A^{(l)}$ stands for the diagonal $(N-1)\times (N-1)$ matrix
obtained from $A$ by removing the eigenvalue $a_l$.
The $c_l$ are normalization constants. The right hand side of $\matpol$ is obviously a
polynomial of degree $N-1$. As the polynomials $Q_l$ are entirely
characterized by property $\orth$, we compute
$$\eqalign{
&\int d\lambda_l Q_l(\lambda_l) e^{N (-V(\lambda_l)+a\lambda_l)}\cr
&=c_l (N-1)! {1\over\Delta(A^{(l)})} \int \prod_{i=1}^N d\lambda_i
\Delta(\lambda_i)_{1\le i\le N} e^{N\sum_{i(\ne l)}(-V(\lambda_i)+a_i \lambda_i)
-NV(\lambda_l)+a \lambda_l}\cr
}\eqn\ortchk$$
where we have introduced the eigenvalues $\lambda_i$, $i\ne l$, of the matrix $M$,
and $\Delta(A^{(l)})\equiv\Delta(a_{l'})_{l'\ne l}$ is the Van der Monde determinant
of the eigenvalues of $A^{(l)}$.
Eq. $\ortchk$ looks like a $N\times N$ matrix integral; if we define $A^{(l),a}$ to
be the diagonal $N\times N$ matrix obtained from $A$ by replacing $a_l$ with $a$,
we have:
$$\int d\lambda_l Q_l(\lambda_l) e^{N (-V(\lambda_l)+a\lambda_l)}
= c_l {1\over N} { \Delta(A^{(l),a})\over \Delta(A^{(l)}) }
\int d^{N^2}\! M \exp(N\tr(-V(M)+MA^{(l),a})).
\eqn\ortchkb$$
If we now set $a=a_{l'}$, $l'\ne l$, the Van der Monde determinant $\Delta(A^{(l),a_{l'}})$
of the eigenvalues of $A^{(l),a_{l'}}$ becomes zero. If $a=a_l$, $A^{(l),a_l}=A$ and
the matrix integral is just the partition function $Z$. For $\orth$ to hold we
need $c_l$ to be:
$$c_l=NZ^{-1}{1\over \prod_{l'(\ne l)} (a_l-a_{l'})}.\eqn\nrm$$

We can finally express the kernel as:
$$\eqalign{
K(\lambda,\mu)&=Z^{-1} e^{-{N\over 2}(V(\lambda)+V(\mu))}
\int \prod_{i=1}^{N-1} d\lambda_i\, e^{-N\sum_{i=1}^{N-1} V(\lambda_i)}\cr
&\Delta(\lambda_i)_{1\le i\le N,\lambda_N\equiv\lambda}
\det(\exp(N\lambda_i a_l))_{1\le i,l\le N,\lambda_N\equiv\mu}.\cr
}\eqn\krn$$
Note that $K$ itself, in contrast with the polynomials $Q_l$, is more
naturally expressed as an integral over eigenvalues than
as a matrix integral.

\section{Saddle point equation and analytic structure}
Before going on with the study of the kernel, we need to understand the analytic
structure of the various functions that we shall now introduce. To do so, we
first write saddle point equations for the standard partition function:
$$Z \sim \int\prod_{i=1}^N d\lambda_i \Delta(\lambda_i)
\det (\exp (N \lambda_j a_l))
e^{-N \sum_{i=1}^N V(\lambda_i)}.\eqn\spz$$
We shall suppose that the density of the $N$ eigenvalues of $A$ has
a smooth limit as $N\rightarrow\infty$. Then the eigenvalues of $M$ also
have a smooth large $N$ limit, characterized by a saddle point distribution:
we shall assume that the eigenvalues fill a single interval $[\alpha,\beta]$
(it will be argued later that this hypothesis is only technical and
does not change the short distance universal behavior),
with a density $\rho(\lambda)\equiv\rho_1(\lambda)$.

Usually, at this stage, one replaces the determinant $\det(\exp(
N\lambda_j a_l))$ with $\exp(N\sum\lambda_i a_i)$ using the symmetry
of exchange of the eigenvalues; here we shall {\it not}
do so, because this would prevent us from writing down a saddle point
equation. Instead, we introduce 2 functions $G(z)$ and $a(z)$ which have a
cut on $[\alpha,\beta]$, such that:
$$\left\{\eqalign{
\Gslash(\lambda_i)&={1\over N} {\der\over\der\lambda_i} \log\Delta(\lambda_i)\cr
\aslash(\lambda_i)&={1\over N} {\der\over\der\lambda_i} \log\det(\exp(N\lambda_i a_l)).\cr
}\right.\eqn\cuts$$
$G$ is of course the resolvent:
$$G(z)=\left<\tr{1\over z-M}\right>.\eqn\reso$$
$a$ cannot be defined by such a simple formula; we refer to appendix 1 for
a rigorous definition of $a$. Here let us note that Itzykson--Zuber formula
$${\det(\exp(N\lambda_i a_l))\over\Delta(\lambda_i)}\sim \int_{\Omega\in U(N)}
d\Omega \, e^{N\tr(\Omega\Lambda\Omega^\dagger A)},\eqn\regiz$$
implies that $\der/\der\lambda_i\log(\det(\exp(N\lambda_i a_l))/\Delta(\lambda_i))$
is a regular function of $\lambda_i$ (i.e. there is no pole when $\lambda_i\sim\lambda_j$);
so we write it under the form $f(\lambda_i)$, where $f$ can be extended 
into an analytic function on the whole complex plane.
Thus $a(z)$ and $G(z)$ are related by:
$$a(z)\equiv G(z)+f(z).\eqn\defa$$
From $\defa$ we deduce that $a$ has the same cut as $G$ on $[\alpha,\beta]$,
that is
$$a(z\pm i0)=\aslash(z)\pm i \pi\rho(z)\qquad \forall z\in [\alpha,\beta]\eqn\samcut$$
$\samcut$ can be taken as a
definition of $a$ on $[\alpha,\beta]$ since $\aslash(z)$ is given by $\cuts$
(see appendix 1 for a definition of $a(z)$ on the whole complex plane and a more detailed
analysis of its properties).

The saddle point equation for $\spz$ is
$$\Gslash(z)+\aslash(z)=V'(z)\qquad\forall z\in [\alpha,\beta]\eqn\spext$$
Using what we know of the analytic structure of $G$ and $a$ (eq. $\samcut$),
we can now extend $\spext$ to the whole complex plane:
$$G(z)+a^{\star}(z)=V'(z).\eqn\wow$$
$a^{\star}$ denotes the function connected to $a$ by the cut $[\alpha,\beta]$.
In other words $a$ is a multi-valued function of $z$,
and $a(z)$ and $a^{\star}(z)$ are the values on opposite sides of the cut
$[\alpha,\beta]$, since (for generic $A$ and $V$) the cut is a square
root-type cut which connects two sheets (figure 1).

\fig{
Analytic structure of $a(z)$ and $G(z)$. In the physical sheet (below), $G(z)$
has a single cut, while in the other sheet (above) it can have more cuts.
It is the opposite for $a(z)$, because of eq. $\wow$}
{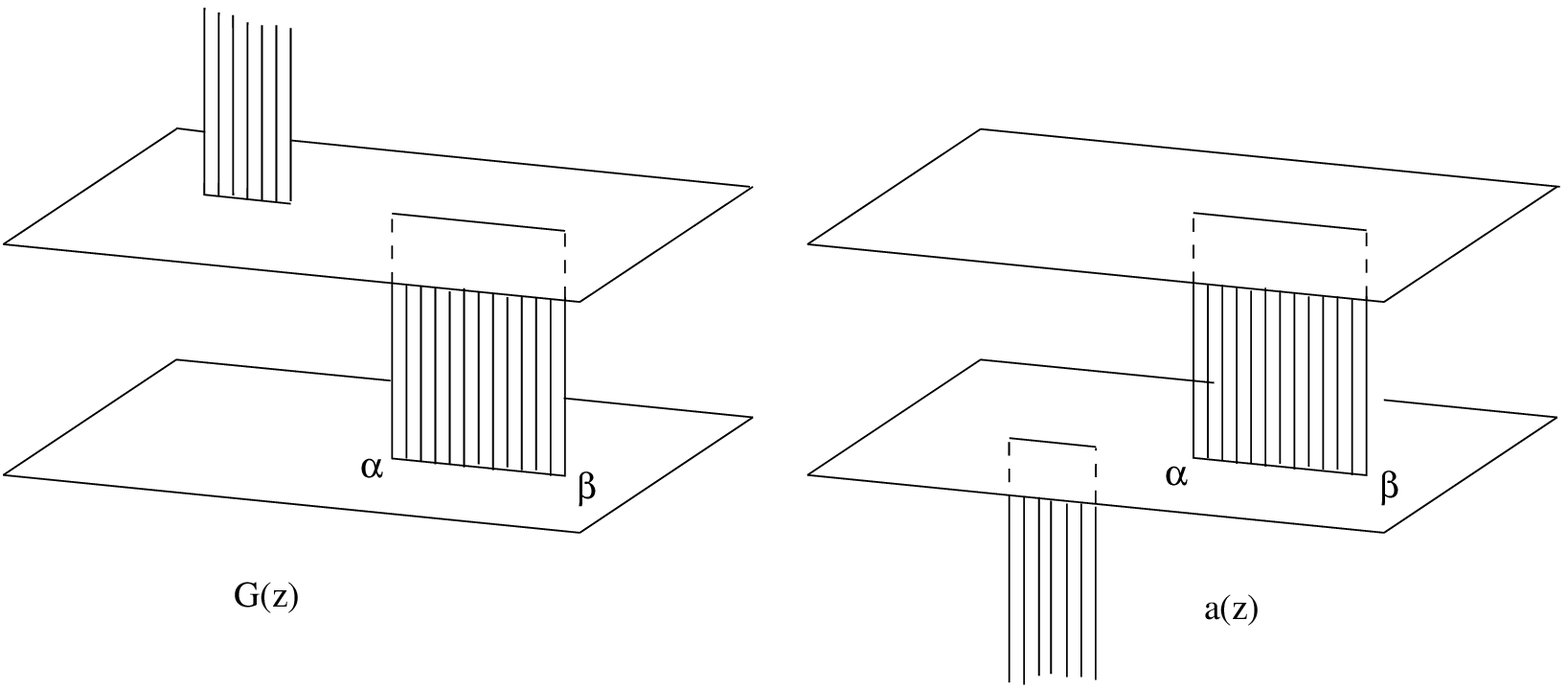}{16cm}

\section{Short distance asymptotics of the kernel.}
Let us analyze the more complicated saddle point equation for the integral
$$Z(\lambda,\mu)=\int \prod_{i=1}^{N-1} d\lambda_i\, e^{-N\sum_{i=1}^{N-1}
V(\lambda_i)}
\Delta(\lambda_i)_{1\le i\le N,\lambda_N\equiv\lambda}
\det(\exp(N\lambda_i a_l))_{1\le i,l\le N,\lambda_N\equiv\mu}\eqn\compz$$
which is related to the kernel $K$ by eq. $\krn$. As in the $U(N)$-invariant case,
$Z(\lambda,\mu)$ can be considered
as the partition function of a model in which the action is the action of $\spz$
(of order $N^2$) plus additional terms dependent on $\lambda$ and $\mu$ (of order $N$).

We shall again write differential equations for $Z$; in a very similar way to
the $U(N)$-invariant case, we find here:
$$\left\{\eqalign{
{\der\over\der\lambda}\log Z(\lambda,\mu)&=(N-1)G_{\lambda,\mu}(\lambda)\cr
{\der\over\der\mu}\log Z(\lambda,\mu)&=(N-1)a_{\lambda,\mu}(\mu).\cr
}\right.\eqn\eqdifx$$
$G_{\lambda,\mu}$ and $a_{\lambda,\mu}$ are defined in the same way as $G$ and $a$
(see appendix 1 for a definition of $a$), but with a modified saddle point distribution
of the $\lambda_i$ due to the additional terms in the action.
Of course, the leading behaviors of $G_{\lambda,\mu}$ and $a_{\lambda,\mu}$,
when $N\rightarrow\infty$ are simply $G$ and $a$, since the corrective terms are
negligible in the leading approximation.

If we wanted to solve the differential equations for
all values of $\lambda$ and $\mu$, we should now calculate the $1/N$ corrections
to the leading behavior. However, as we are only interested in the short distance
behavior of $K$, we can restrict ourselves to the region $\lambda-\mu\sim 1/N$:
variations of $\lambda$, $\mu$ around the diagonal $\lambda=\mu$
(where $K$ is known -- $K(\lambda,\lambda)
=N\rho(\lambda)$) are then of order $1/N$, i.e. we only need the leading behavior
of $\der/\der\lambda \log K$. This means that the next corrections
of $G_{\lambda,\mu}(z)$ and $a_{\lambda,\mu}(z)$
are actually irrelevant in this region, except for possible poles of the type
$1/(z-\lambda)$ or $1/(z-\mu)$, which would be again of order $N$.

Of course $G_{\lambda,\mu}$ and $a_{\lambda,\mu}$ do not have any poles on the
``physical sheet''; but we have learnt from the $U(N)$-invariant case 
that different choices of sheets (or 
of values on the cut joining these sheets, which amounts to the same)
correspond to different saddle points,
and that is how we were led to taking into account saddle points $(\pm,\mp)$ in which
poles at $z=\lambda$ and $z=\mu$ do appear (cf figures 2 and 3 in next section).

We now recall that $a(z)=G(z)+f(z)$, where $f$ is a an analytic function (regular on the
cut $[\alpha,\beta]$). In the same way $a_{\lambda,\mu}(z)=G_{\lambda,\mu}(z)+f_{\lambda,
\mu}(z)$ and it is now clear that poles can only come from the Van der Monde part
of $a(z)$ and not from the regular part $f(z)$. More explicitly, one can write down
a saddle point equation for $G_{\lambda,\mu}$ and $a_{\lambda,\mu}$ which looks like
$$(N-1)\Gslash_{\lambda,\mu}(z)+(N-1)\aslash_{\lambda,\mu}(z)
-NV'(z)+{1\over z-\lambda}+{1\over z-\mu}+\hbox{\rm (regular terms of order $O(N^0)$)} = 0
\eqn\speb$$
At this level of accuracy $f_{\lambda,\mu}=f$, the correction to $f$ being
a regular term of order $1/N$. Then the correction $a_{\lambda,\mu}-a
=G_{\lambda,\mu}-G$ is the same for $a$ or $G$, and
the analysis of eq. $\speb$ becomes perfectly identical
to what was done in section 2 with eq. $\spe$.
We immediately write the differential equations that we obtain:
$$\eqalign{
{\der\over\der\lambda} \log Z_{(\epsilon,\epsilon')} &= \epsilon Ni\pi\rho(\lambda)
+N\Gslash(\lambda)
-{1\over 2} {1\over\lambda-\mu} \left(1-{\epsilon'\sqrt{(\mu-\alpha)(\beta-\mu)}
\over \epsilon \sqrt{(\lambda-\alpha)(\beta-\lambda)}}\right)\cr
{\der\over\der\mu} \log Z_{(\epsilon,\epsilon')} &= \epsilon' Ni\pi\rho(\mu)
+N\aslash(\mu)
-{1\over 2} {1\over\mu-\lambda} \left(1-{\epsilon\sqrt{(\lambda-\alpha)(\beta-\lambda)}
\over \epsilon' \sqrt{(\mu-\alpha)(\beta-\mu)}}\right).\cr
}\eqn\eqdifx$$
Again it is clear that only the saddle-points $(\pm,\mp)$ need to be considered
because the saddle-points $(\pm,\pm)$ are suppressed by a factor of $1/N$.
It is now convenient to introduce a modified kernel; noting that transformations:
$$\tilde{K}(\lambda,\mu)=\gamma(\lambda)K(\lambda,\mu)
{1\over\gamma(\mu)}\eqn\tra$$
do not affect values of determinants of type $\detform$, we choose here
$\gamma(\lambda)=\exp({N\over 2}V(\lambda))$ so that
$$\tilde{K}(\lambda,\mu)\sim e^{-NV(\mu)} Z(\lambda,\mu).\eqn\modker$$
Using the saddle point equation $\spext$, we can rewrite
the differential equations:
$$\left\{\eqalign{
{\der\over\der\lambda} \log \tilde{K}_{\pm,\mp} &=\pm Ni\pi\rho(\lambda)+N\Gslash(\lambda)-
{1\over\lambda-\mu}\cr
{\der\over\der\mu} \log \tilde{K}_{\pm,\mp} &=\mp Ni\pi\rho(\mu)-N\Gslash(\lambda)-
{1\over\mu-\lambda}.\cr
}\right. \eqn\smpdifx$$
Imposing the condition that $\tilde{K}(\lambda,\lambda)=N\rho(\lambda)$, we finally get:
$${\tilde K}(\lambda,\mu)={\sin(N\pi(\lambda-\mu)\rho(\lambda))
\over\pi(\lambda-\mu)}
e^{N(\lambda-\mu) \smallGslash(\lambda)}.\eqn\unix$$
This formula is a generalization of what was obtained in section 2
(which is the case $A=0$). The new factor 
$(\lambda-\mu)\smallGslash(\lambda)$ can be absorbed in a redefinition
of the kernel of the type $\tra$.
\section{Multi-cut generalization.}
So far we have always assumed that the resolvent $G$ has a single
cut, i.e. the saddle point
density of eigenvalues should be non-zero on a single interval $[\alpha,
\beta]$. Intuitively it seems clear that removing the ``single cut''
hypothesis, which is a long distance effect, should
not change the short distance behavior of the kernel.
Indeed, it has been checked in [\DEO]
that the short distance universality is preserved in the quartic 
$U(N)$-invariant case with two symmetric cuts (but the long distance
behavior is modified). Here we shall argue that, in the more general
models we
consider, this change is irrelevant for short distance asymptotics.

Let us start from eq. $\speb$, which was derived without any assumption
on the cuts of $G$ or $a$. We shall study the 
correction $C_\mu(z)$ which satisfies 
$$\Cslash_\mu(z)={1\over 2(z-\mu)}\eqn\mccor$$
and its effect on the differential equations (the same line of reasoning will
apply to $C_\lambda(z)$). Since our interest lies in the region
$\lambda-\mu\sim 1/N$, we can assume that both $\lambda$ and $\mu$ are in
the same interval $[\alpha,\beta]$, which is but one of the cuts of $G$. Of course
$\lambda$ (resp. $\mu$) must be slightly shifted in the imaginary direction
to remove ambiguities, with a shift of $\epsilon i$ (resp. $\epsilon' i$).
As already noted, $C_\mu(z)$ cannot have a pole at $z=\mu$ on the physical sheet.
However,
$$\Cslash_\mu(z)={1\over 2}(C(z+i0)+C^\star(z+i0))={1\over 2}(C(z-i0)+C^\star(z-i0))
\eqn\mcclash$$
where $C^\star(z)$ is the function connected to $C(z)$ by the cut
$[\alpha,\beta]$ (note that the definition of $C^\star$ does not depend on the cut chosen, because of the saddle point equation);
so $\mccor$ can be continued to complex values of $z$ and
one concludes that $C^\star(z)$ has a pole at $z=\mu$, with a residue of $1$.

It is now easy to derive the simplified differential equations for $K_{\epsilon,
\epsilon'}$ in an even more schematic way than before. If $\epsilon=\epsilon'$,
$\lambda$ will never get close to the pole of $C^\star$ (figure 2);
therefore
there will be no $1/(\lambda-\mu)$ term in the different equations and $K_{\epsilon,
\epsilon'}$ will of order $O(N^0)$, i.e. negligible.
On the other hand, if $\epsilon=-\epsilon'$,
$\lambda$ will reach the pole of $C^\star$ (figure 3);
this time we obtain equations $\smpdifx$, and the usual short distance universality.

\fig{\it Analytic structure of $G_{\lambda,\mu}$ and $a_{\lambda,\mu}$ when $\lambda$ and $\mu$ are
on the same side of the cut $[\alpha,\beta]$. When $\lambda\sim\mu$ and the $\lambda$ on the physical sheet 
(indicated by a small arrow) 
approaches the cut, it
does not reach the pole at $z=\mu$ on the other sheet.}
{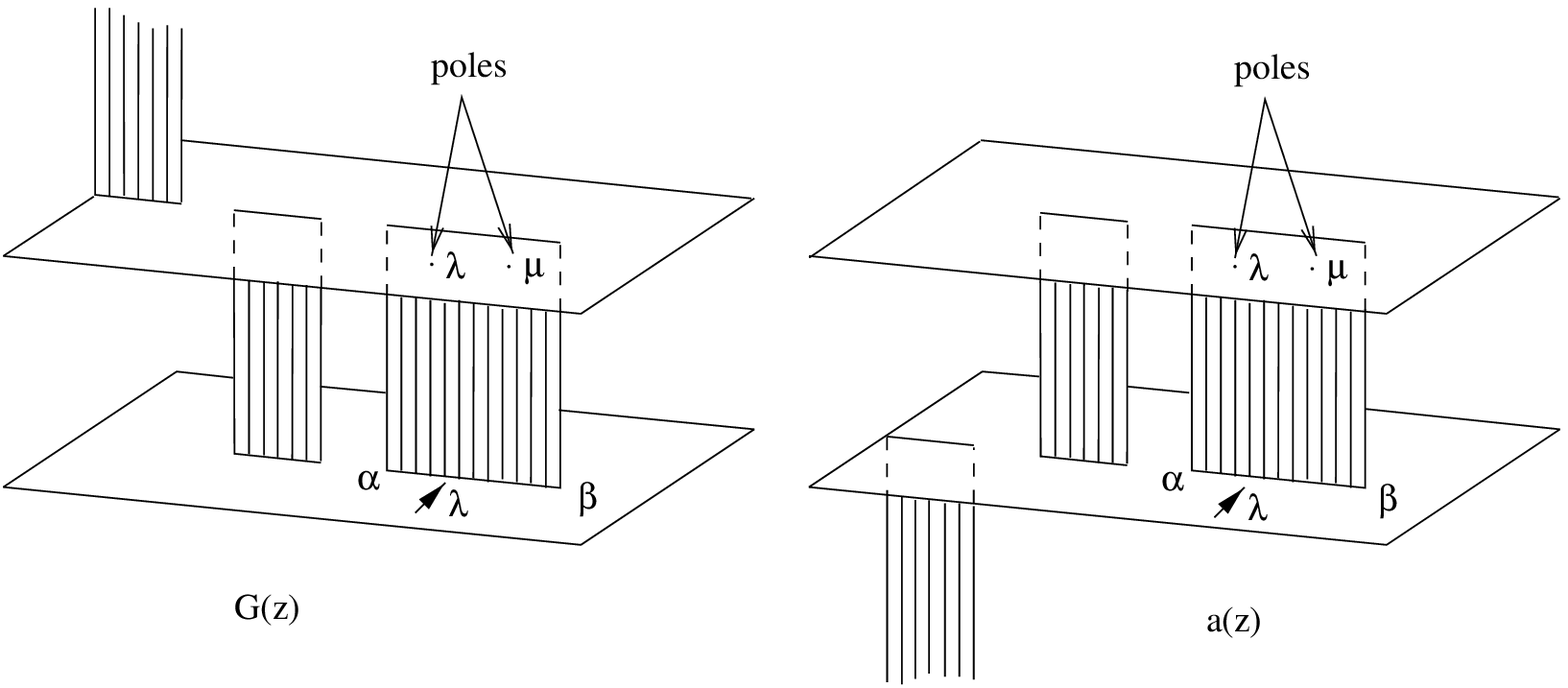}{16cm}

\medskip

\fig{\it Analytic structure of $G_{\lambda,\mu}$ and $a_{\lambda,\mu}$ when $\lambda$ and $\mu$ are
on opposite sides of the cut. This time the $\lambda$ on the physical sheet,
as it approaches the real axis, ``sees'' the pole at $z=\mu$ 
on the other sheet (through the cut).}
{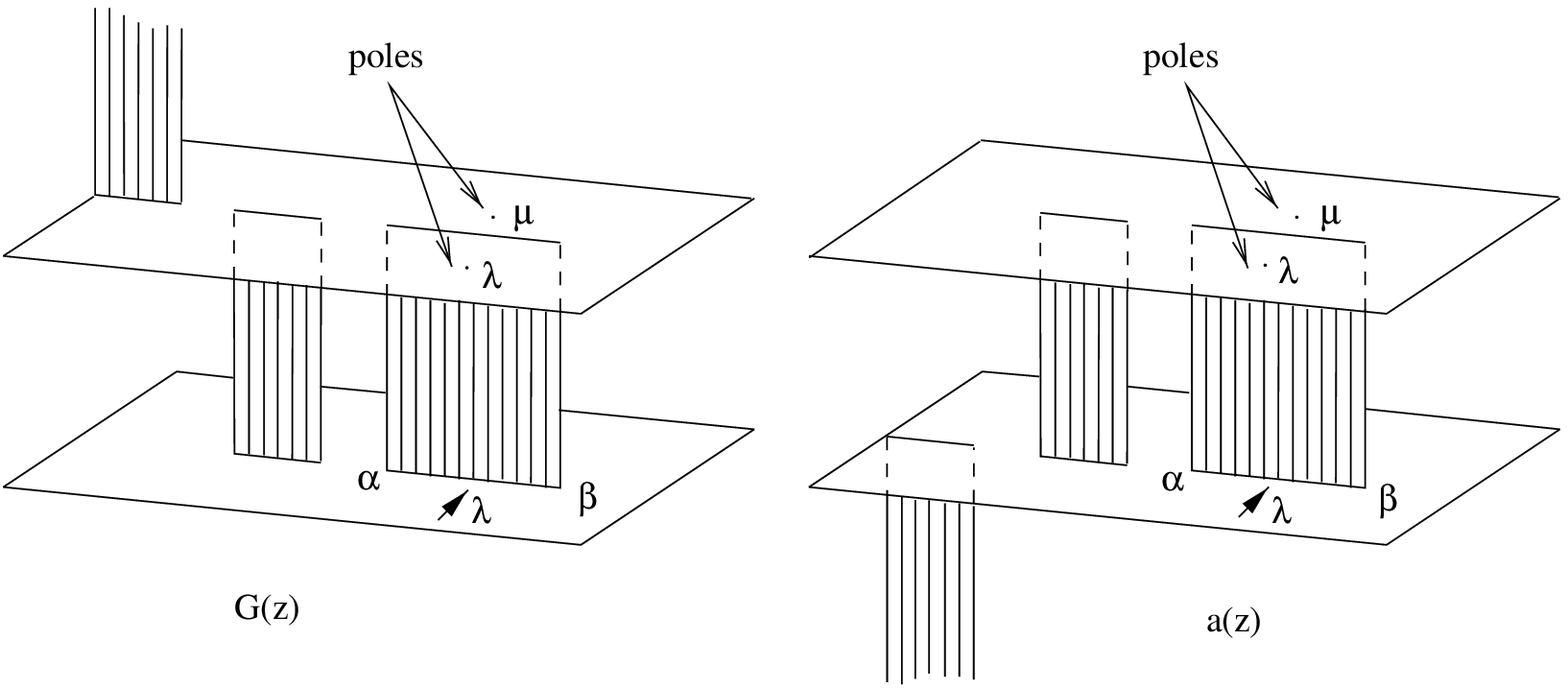}{15cm}

\goodbreak

\chapter{Multi-matrix generalization.}
The results obtained in the previous
sections can be generalized to models of a chain of matrices. We shall only
briefly discuss these, since they are less interesting physically.

\section{The $U(N)$-invariant case.}
The first model considered is defined by the measure (with arbitrary potentials
$V^{(m)}$):
$$\prod_{m=1}^M d^{N^2}M^{(m)} \exp\left(-N\tr \left( -\sum_{m=1}^M V^{(m)}(M^{(m)})
+\sum_{m=1}^{M-1}M^{(m)}M^{(m+1)}\right)\right).\eqn\mulmeas$$
It has a global $U(N)$ invariance, and just as in section 1, can be treated by introducing
the appropriate orthogonal polynomials. Here they are biorthogonal polynomials $P_k$ and $Q_k$
with respect to a non-local measure:
$$\int \prod_{m=1}^M d\lambda^{(m)} P_k(\lambda^{(1)}) Q_l(\lambda^{(M)})
e^{-N\sum_{m=1}^M V^{(m)}(\lambda^{(m)})+N\sum_{m=1}^{M-1}\lambda^{(m)}
\lambda^{(m+1)}}=\delta_{kl}.\eqn\mulpol$$
We can now define the distributions $\rho_n(\lambda_1,\ldots,\lambda_n)$
of the eigenvalues of $M^{(1)}$. It is important to note that
these distributions are different from the
distributions of section 3, after (quenched) averaging over the external field $A$.
They satisfy the usual determinant formulae $\detform$ with the kernel
$$\eqalign{
K(\lambda,\mu)=e^{-{N\over 2}(V(\lambda)+V(\mu))}
&\int \left(\prod_{i=1}^{N-1} d\lambda^{(1)}_i e^{-NV(\lambda^{(1)}_i)} \right)
\left(\prod_{m=2}^M \prod_{i=1}^N d\lambda^{(m)}_i e^{-NV(\lambda^{(m)}_i)}\right)\cr
&\Delta(\lambda^{(1)}_i)_{\lambda^{(1)}_N\equiv\lambda}
\det(\exp(N\lambda^{(1)}_i\lambda^{(2)}_j))_{\lambda^{(1)}_N\equiv\mu}\cr
&\left(\prod_{m=2}^{M-1} \det(\exp(N\lambda^{(m)}_i\lambda^{(m+1)}_j))\right)
\Delta(\lambda^{(M)}_i)
.\cr
}\eqn\mulker$$
We have expressed $K$ directly as an integral over eigenvalues. Note that
at this stage we can replace the determinants 
$\det(\exp(N\lambda^{(m)}_i\lambda^{(m+1)}_j))$
in $\mulker$ with $\exp(N\sum(\lambda^{(m)}_i\lambda^{(m+1)}_i))$, which would
lead to an expression of $K$ in terms of biorthogonal polynomials:
$$K(\lambda,\mu)=e^{-{N\over2}(V(\lambda)+V(\mu))}
\sum_{k=0}^{N-1} P_k(\lambda) \tilde{P}_k(\mu)\eqn\mulkerp$$
where
$$\tilde{P}_k(\mu)=\int \prod_{m=2}^M d\lambda^{(m)} Q_k(\lambda_M)
e^{N\mu\lambda^{(2)}+N\sum_{m=2}^{M-1} \lambda^{(m)}\lambda^{(m+1)}
-N\sum_{m=2}^M V(\lambda^{(m)})}\eqn\mulpolb$$
Here, as in section 3, we do not follow this path: we keep the determinants,
in order to have saddle points. We then write
down differential equations for the kernel, which can be simplified in the
region $\lambda-\mu\sim 1/N$: the resulting equations are identical to
eq. $\smpdifx$, and we find the same asymptotic expression $\unix$.

One could also consider correlations of eigenvalues of a matrix
$M^{(m)}$ somewhere inside the chain, with presumably the same techniques
and results. However, this has not been investigated in detail yet.

\section{The chain of matrices with an external field at one end.}
In this model we add to the measure $\mulmeas$ 
of the preceding section an external
source term for $M^{(1)}$:
$$\prod_{m=1}^M d^{N^2}M^{(m)} \exp\left(-N\tr \left( -\sum_{m=1}^M V^{(m)}(M^{(m)})
+AM^{(1)}+\sum_{m=1}^{M-1}M^{(m)}M^{(m+1)}\right)\right).\eqn\mulmeasx$$
This measure is no more $U(N)$-invariant. We define again the
distributions of eigenvalues of $M^{(1)}$, which satisfy determinant
formulae $\detform$: we give without proof the kernel
$$\eqalign{
K(\lambda,\mu)=e^{-{N\over 2}(V(\lambda)+V(\mu))}
&\int \left(\prod_{i=1}^{N-1} d\lambda^{(1)}_i e^{-NV(\lambda^{(1)}_i)} \right)
\left(\prod_{m=2}^M \prod_{i=1}^N d\lambda^{(m)}_i e^{-NV(\lambda^{(m)}_i)}\right)\cr
&\det(\exp(Na_i\lambda^{(1)}_j))_{\lambda^{(1)}_N\equiv\lambda}
\det(\exp(N\lambda^{(1)}_i\lambda^{(2)}_j))_{\lambda^{(1)}_N\equiv\mu}\cr
&\left(\prod_{m=2}^{M-1} \det(\exp(N\lambda^{(m)}_i\lambda^{(m+1)}_j))\right)
\Delta(\lambda^{(M)}_i)
.\cr
}\eqn\mulkerx$$
The differential equations will look a little more general; for example
the equivalent of eq. $\eqdifx$ will be:
$$\left\{\eqalign{
{\der\over\der\lambda}\log Z(\lambda,\mu)&=(N-1)a_{\lambda,\mu}(\lambda)\cr
{\der\over\der\mu}\log Z(\lambda,\mu)&=(N-1)\lambda^{(2)}_{\lambda,\mu}(\mu).\cr
}\right.\eqn\muleqdifx$$
where 2 functions $a(\lambda^{(1)})$ and $\lambda^{(2)}(\lambda^{(1)})$
 must be introduced
(the logarithmic derivatives of $\det(\exp(Na_i\lambda^{(1)}_j))$
and $\det(\exp(N\lambda^{(1)}_i\lambda^{(2)}_j))$). 
The rest of the analysis is the same. At the end, one redefines
the kernel $K$ with a transformation of type $\tra$, using 
the saddle point equation:
$$\aslash(z)+\lambdaslash^{(2)}(z)=V'(z)$$
so that it satisfies the universal property $\asymp$.

\vfill\eject

\chapter{Conclusion.}
Let us summarize our results. The model considered is that
of a matrix coupled to an external field $A$; the latter has a smooth large $N$
limit characterized by a limiting density of eigenvalues. A kernel
$K(\lambda,\mu)$ is defined such that the determinant formulae $\detform$
hold. In the simplest case where $A=0$, the large $N$ form of $K(\lambda,\mu)$
for all $\lambda$ and $\mu$, first found in [\BZ], is reproduced here.
In the case of a non-zero $A$, we restrict ourselves to the region
$\lambda-\mu\sim 1/N$ (it is not clear that, for general $A$ and $V$,
the long distance behavior of $K$, even after smoothing the oscillations,
should be interesting, e.g. should exhibit any kind of universal behavior).
The asymptotic form $\unix$ is obtained, extending
the level spacing universality to this class of models.

The key ingredient of the derivation of the short distance universality
is of course the existence of the kernel.
But it seems reasonable to assume that
the level-spacing universality (observed experimentally for a broad range
of systems)
should be true for very general matrix models,
in which the correlation functions do not satisfy $\detform$. The problem
is then to manage to compute the level-spacing distribution $P(s)$, even
though it is no more simply related (through $K$, cf eq. $\lvlspc$) to
the correlation functions, that is the naturally
calculable quantities of the model. This suggests that a totally different
approach is probably necessary. Here, let us mention that the question 
of knowing how far the universality of $P(s)$ can be extended is
reminiscent of the question of knowing what is the domain of attraction
of a fixed point in renormalization group theory. What we have shown is that in
our matrix models, both non-gaussian
terms and terms explicitly breaking the $U(N)$-invariance are
irrelevant in the large $N$ limit and lead to the gaussian $U(N)$-invariant fixed point.
Maybe RG methods can
be applied here too (cf [\BZJ]) and allow a much more general approach to
the problem.

In the case of multi-matrix models, the same short distance behavior
is found for the correlation functions of the
eigenvalues of a given matrix, here the first matrix
in a chain of matrices. Of course, it would be interesting to investigate
the more general problem of the correlations between eigenvalues of different
matrices in the chain. This has been already done
in the gaussian case [\BH]. The
conclusion is that the interesting limit is that of an infinite chain,
which tends to the $c=1$ matrix model
$$Z=\int [dM(t)] \exp\left[-N \tr(V(M(t),t)+\dot{M}^2) \right].$$
One should then study the correlations of the eigenvalues of $M(t_1)$ and
$M(t_2)$ as $t_1$ and $t_2$ are close to each other. In order
to apply the methods used in this paper, one will of course
have to generalize the determinant formulae $\detform$ to include eigenvalues
of different matrices.

\medskip
\centerline{\bf Acknowledgment.}
I would like to thank E.~Br\'ezin and B.~Eynard
for stimulating discussions.

\bigskip

\APPENDIX{A1}{1. Analytic structure and functional inverses.}
Let us define the function $a(\lambda)$ used in section 3,
and its functional inverse
$\lambda(a)$. We shall assume, as in [\MAT], that the logarithm of the
Itzykson--Zuber integral
$$W={1\over N^2} \log {\det(\exp(Na_l\lambda_i))\over\Delta(\lambda_i)\Delta(a_l)}\eqn\w$$
has a smooth large $N$ limit, so that it depends only on the distributions of eigenvalues
$\rho(\lambda)$ and $\tilde{\rho}(a)$ of $M$ and $A$. Then one can define functional
derivatives of $W$ with respect to $\rho$ and $\tilde{\rho}$. It is clear (eq. $\cuts$)
that
$$\aslash(\lambda)-\Gslash(\lambda)
={d\over d\lambda}{\delta\over\delta\rho(\lambda)} W[\rho,\tilde{\rho}].\eqn\dw$$
The r. h. s. can be extended to complex values of $\lambda$ and has no cut
on $[\alpha,\beta]$ (the support of the density $\rho$), therefore we can define
$a(\lambda)$ by removing the slashes in $\dw$. Note that this corresponds to the
very simple finite $N$ definition:
$$a(\lambda)\equiv {1\over N}{d\over d\lambda} 
\log\det(\exp(Na_l\lambda_i))_{\lambda_{N+1}\equiv\lambda}\eqn\smpdef$$
The only problem of this definition is that there are now $N$ eigenvalues $a_l$ and $N+1$
eigenvalues $\lambda_i$: one should add one $a_l$ or remove one $\lambda_i$.
In the large $N$ limit, this problems disappears,
since if we believe that $W$ depends only on $\rho(\lambda)$ and $\tilde{\rho}
(a)$, then one can redefine the eigenvalues $a_l$ (or $\lambda_i$) to add an eigenvalue
$a_l$ (or remove one $\lambda_i$), keeping the densities $\tilde{\rho}(a)$ and $\rho(\lambda)$
fixed.

All that has been done so far is symmetric in the exchange of $A$ and $M$, so we can
also define $\lambda(a)$:
$$\lambda(a)=\tilde{G}(a)+{d\over da}{\delta\over\delta\tilde{\rho}(a)} W[\rho,\tilde{\rho}].
\eqn\dwb$$
where $\tilde{G}(a)$ is the resolvent of $A$, and
the functional derivative has been again extended to complex values of $a$.

Let us now discuss the analytical structure of $a(\lambda)$ and $\lambda(a)$. On $[\alpha,
\beta]$, according to $\dw$, $a(\lambda)$ has the same cut as $G(\lambda)$, i.e.
$$\Im a(\lambda\pm i0)=\pm\pi\rho(\lambda).\eqn\ima$$
Likewise, $\lambda(a)$ has the same cut as $\tilde{G}(a)$. We can then define
$a^\star(\lambda)$ (resp. $\lambda^\star(a)$), to be the functions on the other side
of the cut of $G$ (resp . $\tilde{G}$).
If we now consider the matrix model with an external source term (measure $\measb$),
$a^\star(\lambda)$ satisfies an additional constraint which is the saddle point equation $\wow$.
One can see that this implies that $a^\star(\lambda)$, just like
$G(\lambda)$, has no other cut in the whole complex plane than
that on $[\alpha,\beta]$.
On the other hand, $\lambda^\star(a)$ is not constrained and therefore it may have more cuts
(leading to other sheets).

Finally, it can be shown by studying the large $N$ limit of the Itzykson--Zuber integral
that $\lambda(a)$ and $a(\lambda)$ (as multi-valued functions) are functional inverses of each
other. This can be thought of as a generalization of the inversion
relation found in [\MAT], even though the connection is non-trivial.
Here we shall derive this relation in an elementary fashion.

We rewrite definition $\smpdef$ explicitly:
$$a(\lambda)={\sum_{l=1}^{N+1} a_l\, E_{l,N+1} e^{Na_l \lambda}
\over \sum_{l=1}^{N+1} E_{l,N+1} e^{Na_l\lambda}}\eqn\expdef$$
where $E_{l,N+1}$ is the determinant $\det(\exp(Na_k\lambda_i))$ with
$1\le k\le N+1$, $k\ne l$ and $1\le i\le N$. Now eq. $\w$ can be applied
to $E_{l,N+1}$: 
$${1\over N^2}\log {E_{l,N+1}\over\Delta(\lambda_i)\Delta(a_k)_{k\ne l}}
=W[\rho,\tilde{\rho}_l]+O(1/N^2)\eqn\wl$$
where $\tilde{\rho}_l$ is the density of the $a_k$, $k\ne l$;
computing the $1/N$ correction to $\log E_{l,N+1}$ one finds
$$\log E_{l,N+1}
=C-N {\delta\over\delta\tilde{\rho}(a_l)}W[\rho,\tilde{\rho}]
-\sum_{k(\ne l)} \log(a_k-a_l)\eqn\eln$$
where $C$ is independent of $l$. Finally, using the standard trick
which is to replace the sum over $l$ with a contour integral,
$\expdef$ becomes
$$a(\lambda)
={\oint da \tilde{G}(a) \, a\, e^{Na\lambda}e^{-N\int^a \lambda(a')da'}
\over\oint da \tilde{G}(a) \, e^{Na\lambda}e^{-N\int^a \lambda(a')da'}
}\eqn\intdef$$
The saddle point equation gives simply: $a(\lambda)=a$ with
$\lambda(a)=\lambda$, which proves the functional inversion relation.

A last remark: the analytic structure described here can be made more explicit
in the gaussian case (with an external field), by using Pastur's
self-consistent relation [\PA].

\APPENDIX{A2}{2. Connection with characters.}
One may notice the strong resemblance between
equations discussed in section 3.3 and appendix 1 and large $N$ character relations
found by Kazakov et al. in [\KSW]. To establish the connection, we shall now rederive
in a very simple manner the latter relations.

A representation of $U(N)$ can be described
by its highest weights $m_1\ge m_2 \ge \ldots \ge m_N$.
The corresponding character is defined by:
$$\chi_h(M)={\det(z_i^{h_j})\over\Delta(z_i)}\eqn\defcar$$
where the $z_i$ are eigenvalues of the $N\times N$ matrix $M$, and the $h_i$ are
the shifted weights $h_i=m_i+N-i$.

If the weights are positive ($m_N\ge 0$, which we shall simply write $h\ge 0$)
then we can use the identity
$$\sum_{h'\ge 0} \chi_{h'}(M) \chi_{h'}(M_0)=\exp\left(\sum_{q=1}^{\infty} {1\over q}
\tr M^q \tr M_0^q \right)\eqn\idt$$
and the orthogonality of characters to compute
$$\eqalign{
\chi_h(M_0)&=\int_{M\in U(N)} dM
\bar{\chi}_h(M)
\left(\sum_{h'\ge 0} \chi_{h'}(M) \chi_{h'}(M_0)\right)\cr
&=\int_{M\in U(N)} dM \bar{\chi}_h(M)
\exp(N\tr V(M))\cr
}\eqn\cpt$$
The potential $V(z)=\sum_{q=1}^{\infty} t_q z^q/q$
is related to $M_0$ by $t_q={1\over N}\tr
(M_0^q)$.

We now diagonalize $M$ and integrate over the unitary group;
we are left with the eigenvalues
$z_i$ of $M$, $|z_i|=1$; using the definition $\defcar$ we obtain:
$$\chi_h(M_0)=\oint\prod_{i=0}^{N-1} {dz_i\over z_i}
\Delta(z_i) \det(z_i^{-Nh_j}) e^{N\sum_{i=0}^{N-1} V(z_i)}.\eqn\pfc$$
(the shifted weights $h_j$ have been rescaled by a factor of $N$).
This is to be compared with partition function $\spz$. The study
of the two cases (and in particular the analytic structure
that arises) being very similar,
we shall now skip the details of the derivation.

One first writes down a saddle point equation:
$$\kslash(z)+\Gslash(z)+V'(z)=0\eqn\spec$$
where the function $k(z)$ is defined by:
$$\kslash(z_i)={1\over N} {\der\over\der z_i} \log\det(z_i^{-Nh_j}).\eqn\kdef$$
$k(z)$ has the same cut as $G(z)$, i.e. where the saddle point density of eigenvalues
is non-zero. We then have
$$k^{\star}(z)+G(z)=V'(z)\eqn\specb$$
on the whole complex plane. ($k^{\star}$ is the function on
the other side of the cut of $k$).
We make the trivial redefinition $h(z)=-zk^{\star}(z)$, and
expand in powers of z:
$$\eqalign{
h(z)&=\sum_{q=1}^{\infty} t_q z^q+1+
\sum_{q=1}^{\infty} z^{-q} \left({1\over N}\sum_{i=0}^{N-1} z_i^q\right)\cr
&=\sum_{q=1}^{\infty} t_q z^q+1+
\sum_{q=1}^{\infty} z^{-q} {1\over N^2} q {\der\over \der t_q} \log\chi_h(M_0)\cr
}\eqn\ksw$$
which is precisely the result obtained by much more complicated
methods in [\KSW] (in their notation $z\equiv G^{-1}$.)

Finally, the same functional inversion argument as in appendix 1
applies here, proving that $z(h)$ defined by $z(h)=\exp w(h)$ with
$$\wslash(h_i)=-{1\over N}{\der\over\der h_i} \log\chi_h(M_0)\eqn\wdef$$
(cf [\KSW] for a similar definition of $G(h)$)
is the functional inverse of $h(z)$.

\vfill\eject

\refout
\end